\documentclass[aoas,MSNbibl,nameyear,seceqn,dvips]{arximspdf}
\usepackage{graphicx}
\doi{10.1214/13-AOAS657} 
\volume{7}
\issue{3}
\pubyear{2013}
\firstpage{1814}
\lastpage{1835}

\makeatletter

\newcommand{\rright}{\right}
\newcommand{\lleft}{\left}
\def\cal{\mathcal}
\makeatother

\begin{document}
\begin{frontmatter}

\title{A decision-theoretic approach for segmental classification}
\runtitle{Decision-theoretic approach for segmental classification}

\begin{aug}
\author[a]{\fnms{Christopher} \snm{Yau}\corref{}\thanksref{t2}\ead[label=e1]{c.yau@imperial.ac.uk}}
\and
\author[b]{\fnms{Christopher C.} \snm{Holmes}\thanksref{t3}\ead[label=e2]{cholmes@stats.ox.ac.uk}}
\thankstext{t2}{Supported in part by a UK Engineering and Physical
Sciences Research Council Life Sciences
Interface Doctoral Training Studentship and by a UK Medical Research
Council Specialist Training Fellowship in
Biomedical Informatics (Ref No. G0701810).}
\thankstext{t3}{Supported in part by a UK Medical Research Council
Programme Leaders Award.}
\runauthor{C. Yau and C.~C. Holmes}
\affiliation{Imperial College London and University of
Oxford}
\address[a]{Department of Mathematics\\
Imperial College London\\
South Kensington Campus\\
London SW7 2AZ\\
United Kingdom\\
\printead{e1}}
\address[b]{Department of Statistics\\
University of Oxford\\
1 South Parks Road\\
Oxford OX1 3TG\\
United Kingdom\\
\printead{e2}}
\end{aug}

\received{\smonth{7} \syear{2010}}
\revised{\smonth{5} \syear{2013}}

%
\begin{abstract}
This paper is concerned with statistical methods for the segmental
classification of linear sequence data where the task is to segment and
classify the data according to an underlying hidden discrete state
sequence. Such analysis is commonplace in the empirical sciences
including genomics, finance and speech processing. In particular, we
are interested in answering the following question: given data $y$ and
a statistical model $\pi(x,y)$ of the hidden states~$x$, what should we
report as the prediction $\hat{x}$ under the posterior distribution $\pi
(x|y)$? That is, how should you make a prediction of the underlying
states? We demonstrate that traditional approaches such as reporting
the most probable state sequence or most probable set of marginal
predictions can give undesirable classification artefacts and offer
limited control over the properties of the prediction. We propose a
decision theoretic approach using a novel class of Markov loss
functions and report $\hat{x}$ via the principle of minimum expected
loss (maximum expected utility). We demonstrate that the sequence of
minimum expected loss under the Markov loss function can be enumerated
exactly using dynamic programming methods and that it offers
flexibility and performance improvements over existing techniques. The
result is generic and applicable to any probabilistic model on a
sequence, such as Hidden Markov models, change point or product
partition models.
\end{abstract}

%

\end{frontmatter}

\section{Introduction}

This paper is concerned with statistical methods for the segmental
analysis of linear sequence data where the task is to segment and
classify data according to an unobserved discrete state sequence. Such
analysis is commonplace in the empirical sciences including genomics
[\citet{Day2007,Majoros2004b,Su2008}], finance [\citet
{Chopin2004,Crowder2005,Rossi2006,Banachewicz2008}] and speech
processing [\citet{Chien2005,Yan2007,Weiss2008}]. In particular, we are
interested in answering the question: given data $y$ and a statistical
model $\pi(x,y)$ of the hidden states $x$, what shall we report as the
prediction $\hat{x}$?

In this paper we formalise the segmental classification problem within
a Bayesian decision theoretic framework. We propose a new class of
Markov loss function that penalises the misclassification of state
occupancy \emph{and} transitions which are errors of direct relevance
in many segmental classification problems. Under the Markov loss
function, the state sequence with minimum expected loss (or maximum
expected utility) can be enumerated using dynamic programming methods
and can provide a simple, yet effective, means of reporting for many
pre-existing statistical models of linear sequence data.

Note that throughout we will make a clear distinction between the \emph
{modeling} task, which involves designing and fitting the best possible
statistical model for $\pi(x,y)$, and the \emph{prediction} task, that
we address here, which involves finding a procedure to obtain a
segmental prediction upon which actions are taken.

\section{Application}

Our motivating application is the problem of identifying DNA copy
number alterations from modern high-throughput genomic technologies:
array comparative genomic hybridisation (aCGH), single nucleotide
polymorphism (SNP) genotyping data or next generation sequencing (NGS).
Copy number alterations are segments of DNA that occur at variable copy
number relative to a reference genome. In humans, we typically possess
two copies of every gene, one inherited from each of our parents.
However, in genomic regions containing copy number alterations, it is
possible to have less than two copies, in which case that region is
said to harbour a copy number loss or \emph{deletion}, or more than two
copies, where the region is then said to contain a \emph{duplication}.
In rare genetic disorders, whole or partial copies of entire
chromosomes can be lost or gained; for example, Downs Syndrome is
caused by the gain of an extra copy of chromosome~21. Our particular
interest lies in copy number profiling of genomically complex cancers
where copy number alterations can arise due to mutations that disrupt
the normal function of DNA repair and chromosome segregation during
cell division.

\begin{figure}

\includegraphics{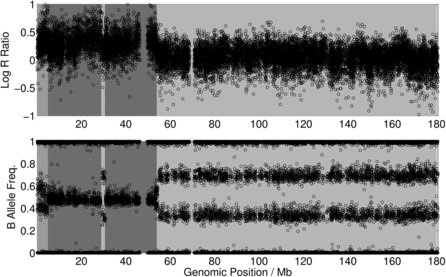}

\caption{Example: SNP genotyping data. A SNP genotyping data comprises
two sets of measurements---the Log R Ratio and the B allele frequency-measured
at multiple locations along the genome. Alterations in the
distributions of the measurements correspond to underlying changes in
the DNA copy number. Each
coloured region corresponds to a different underlying DNA copy number state.}
\label{fig:example_data}
\end{figure}

As an illustration, Figure~\ref{fig:example_data} depicts a SNP
genotyping data set that measures variation in DNA copy number along a
particular chromosome from DNA derived from tumour cells. The
statistical problem is to divide the sequence into regions and to
classify each region by the underlying DNA copy number. This task is
typically made substantially more challenging in cancer due to
confounding factors such as aneuploidy, intra-tumour heterogeneity and
normal cell contamination. These issues are reviewed and discussed in
\citet{Loo2012a}.\vadjust{\goodbreak} Genome-wide profiling of copy number alterations in
cancers [\citet
{Bignell2010,Beroukhim2010a,Curtis2012,Northcott2012,Knight2012}] have
typically employed the use of a variety of statistical approaches for
generating copy number profiles [\citet
{Popova2009,Loo2010,Greenman2010,Yau2010,Li2011,Carter2012}].

A popular class of methods is based on the use of Hidden Markov models
where the hidden state is used to denote the unknown copy number at a
particular location. Copy number sequence predictions are then reported
by finding the most probable state sequence using the Viterbi algorithm
or the most probable set of marginal predictions using the
forward--backward algorithm. A potential limitation of discrete models,
such as the HMM, for cancer analysis is the possibility of cellular
heterogeneity in tumour samples. This can be problematic for aCGH data,
as differences in signal intensity level may correspond to cell-to-cell
variation rather than actual copy number changes. With SNP arrays the
availability of allele-specific intensity data can mitigate the
problem. Statistical models [\citet
{Popova2009,Loo2010,Yau2010,Li2011,Carter2012}] have been developed
that modeled the structure of allele-specific signals that results from
certain types of cellular heterogeneity.

With modern high-density microarrays and next generation sequencing
data it is possible to reveal many hundreds of structural aberrations
within a single tumour. These aberrations can range in size from large,
whole or partial chromosomal gains and losses to small focal
aberrations affecting potential driver mutations (oncogenes and tumour
suppressors). Current state-of-the-art methods can report accurate copy
profiles but can lead to practical problems: a collection of lengthy,
unmanageable lists of genomic alterations that must be screened by
cancer biologists. In this paper, we will show that our
decision-theoretic methods can be used to augment existing models and
provide increased flexibility for sequence classification. We
demonstrate the utility of these methods as a means to report \emph
{smoother} copy number profiles that retain key copy number alterations
while having reduced overall complexity.

\section{Motivation}

\subsection{Decision theory}
We begin by defining some notation. Let $x_i \in\{0,\break\ldots,S\}$ denote
the true unobserved underlying state at the $i = 1, \ldots, n$
locations, and $y_i$ the corresponding observation. The task is to
obtain a prediction $\hat{x} = \{\hat{x}_1, \ldots, \hat{x}_n\}$ given
a statistical model $\pi(x|y)$ [for notational simplicity, we shall
suppress the conditioning on $y$ in the following and refer to $\pi
(x|y)$ as $\pi(x)$].

Bayesian decision theory [\citet{Berger85, Bernardo2000}] provides an
axiomatic framework for making optimal decisions via the principle of
minimum expected loss (or maximum expected utility). In our problem the
``decision'' is the reporting of $\hat{x}$ from which a set of actions
will be taken with associated losses based on the unknown true state of
nature $x$. We encapsulate the forms of error into a loss function
$l(\hat{x} | x)$ which quantifies the loss of taking actions with $\hat
{x}$ when the true state of nature is $x$. The principle of minimum
expected loss (MEL) prescribes one should report $\hat{x}$ as
\begin{eqnarray*}
\hat{x} & = & \arg\min_{\tilde{x}} \mathbb{E}_{\pi(x)} \bigl[
l(\tilde{x} | x) \bigr],
\\
& = & \arg\min_{\tilde{x}} \sum_{x} l(
\tilde{x} | x) \pi(x).
\end{eqnarray*}

\subsection{Standard summaries for segmental classification}

Two summary predictions that are often used for $\hat{x}$ are as
follows: (i) the most probable sequence $\hat{x} = \arg\max_x \pi(x)$
(MAP) or (ii) the set of marginally most probable classifications
(MaxMarg), $\hat{x}_i = \arg\max_{x_i} \sum_{x_{-i}} \pi(\{x_i,
x_{-i}\})$ where the summation is over $x_{-i}$, the state sequence
other than $x_i$. From a decision theoretic perspective, it is
interesting to note the corresponding loss functions that would
motivate the use of these summaries.

In the case of the MAP sequence, the implicit loss function is the following:
%
\begin{equation}
l_G(\hat{x} | x) = \cases{ %
0, &\quad ${\mbox{if }} \hat{x} \equiv x,$
\vspace*{2pt}\cr
1, &\quad ${\mbox{otherwise}}.$}
\label{eq:global_loss}
\end{equation}
We shall refer to this as the \emph{global loss} function, as a
constant penalty is incurred if the prediction is not completely
correct. This loss function is extreme in the sense that no matter how
many misclassification errors are made, the same\vadjust{\goodbreak} penalty is incurred,
that is, it is an ``all-or-nothing approach.'' Furthermore, for this loss
function the entirety of the sequence is important, the optimal
prediction must be globally and locally correct.

\begin{figure}[b]

\includegraphics{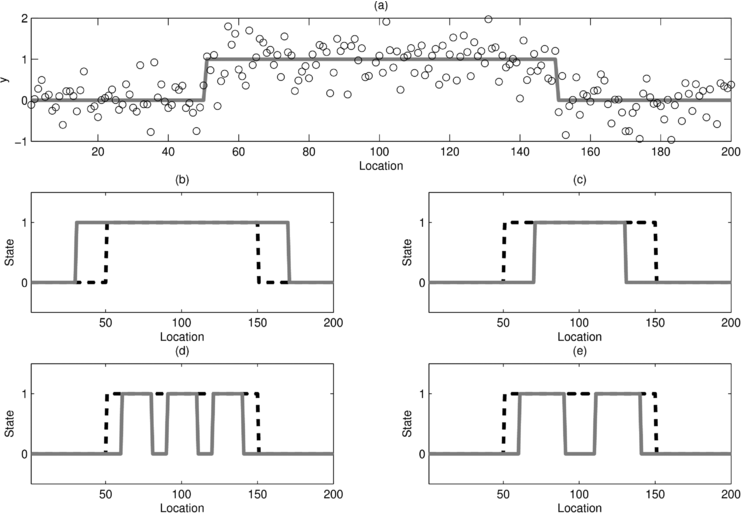}

\caption{Sequence predictions. An example data set \textup{(a)} and four
predictions of the underlying state sequence \textup{(b)--(e)}. (Grey, solid)
Predicted and (Black, dashed) true state sequence.}
\label{fig:example_decisions}
\end{figure}

For the MaxMarg sequence, the implicit loss function assumed is as follows:
\[
l(\hat{x} | x) = \sum_{i} l_M(
\hat{x}_i | x_i),
\]
with
%
\begin{equation}
l_M(\hat{x}_i | x_i) = \cases{
0, &\quad ${\mbox{if }} \hat{x}_i \equiv
x_i,$
\vspace*{2pt}\cr
\mathrm{FC}, &\quad ${\mbox{otherwise}},$}
\label{eq:marginal_loss}
\end{equation}
where $\mathrm{FC}$ is the cost of making a false classification. We
shall refer to this as the \emph{marginal loss} function.

In contrast to the global loss function, the marginal loss function
ignores any form of local or global structure. It concentrates instead
on penalising classification error at each location considered
independently of others, which is equivalent to stating that the
overall loss is invariant to permutations of the sequence $\{\hat{x}_i,
x_i\}_{i=1}^n$. As a result, if we consider the simulated data sequence
in Figure~\ref{fig:example_decisions}(a)\vadjust{\goodbreak} which contains a region of
elevated signal related to an underlying change in the hidden state,
the predictions shown in Figure~\ref{fig:example_decisions}(b)--(e) which
contain the same number of misclassifications may incur the same loss
under the marginal loss function even though each prediction is
qualitatively very different and may contain a different number of
predicted segments that could lead to quite different actions if
decisions are taken upon them. It is clear, therefore, that simply
counting the number of state misclassifications is insufficient.

\subsection{Limitations of standard summaries}

These two commonly used loss functions correspond to quite opposite
extremes and neither scenario seems appropriate in segmental
classification problems. For example, in many situations it is unusual
for classification errors to be completely intolerable, instead there
are acceptable tolerance levels for error. Under these circumstances it
would not be appropriate to use the global loss function in which the
same penalty is incurred irrespective of how many errors are made in
the prediction. Moreover, there is no flexibility with the global loss
and the user cannot explore other predictions with fewer or greater
number of transitions. Furthermore, if we are interested in segmental
classification and we expect dependencies between states at different
locations, it does not seem appropriate to use a marginal loss function
that considers classification error at each location independently of
the others.

Nonetheless, the appeal of these loss functions is that the computation
of the state sequence with minimum expected loss is often analytically
tractable or simple to approximate with commonly used statistical
models. For example, in Hidden Markov models, the Viterbi algorithm
allows the most probable sequence to be enumerated exactly while the
forward--backward algorithm allows the marginal probabilities $\pi(x_i)
= \sum_{x_{-i}} \pi( x )$ with computational time complexity that is
linear in the length of the data sequence [\citet{Rabiner89}].


\section{Method}

\subsection{Markov loss function}

We now introduce a loss function for segmental classification that
penalises incorrect state classifications and transitions:
\[
l_{\mathrm{ML}}(\tilde{x} | x) = \sum_{i=1}^{n}
l_M(\tilde {x}_{i}|x_{i}) + \sum
_{i=1}^{n-1} l_T(\tilde{x}_{i,i+1}|x_{i,i+1})
,
\]
where $x_{i,i+1}$ denotes the pair $\{x_i, x_{i+1}\}$. We refer to this
as the \textit{Markov loss} function. This loss function extends the
marginal loss function $l_M(\tilde{x}|x)$ to include penalty terms on
state transition errors $l_T(\tilde{x}_{i,i+1}|x_{i,i+1})$ as follows:
\begin{eqnarray*}
l_T(\tilde{x}_{i,i+1}|x_{i,i+1}) = \cases{
\mathrm{FT}, & \quad$\mbox{if } \tilde{x}_{i} \neq\tilde{x}_{i+1},
x_i = x_{i+1}$, \vspace*{2pt}
\cr
\mathrm{FH}, &\quad $
\mbox{if } \tilde{x}_{i} = \tilde{x}_{i+1}, x_i
\neq x_{i+1},$\vspace*{2pt}
\cr
0, & \quad $\mathrm{otherwise},$}
\end{eqnarray*}
where for exposition we assume a common cost of error irrespective of
the actual state.

The Markov loss function contains three parameters: (i) $\mathrm{FC}$
(False Call)---cost of a state classification error, (ii) $\mathrm{FT}$
(False Transition)---the cost associated with calling a false state
transition and (iii) $\mathrm{FH}$ (False Hold)---the cost of
incorrectly staying in the same state. In the special case when $\mathrm
{FT}=\mathrm{FH}=0$, the Markov loss function reduces to the marginal
loss function which forms a subclass of our more general loss function.
An example pairwise loss function for a binary state problem is shown
in Table~\ref{tab:cost_matrix}.

\begin{table}
\caption{Cost matrix structure for binary state transition}\label{tab:cost_matrix}
\begin{tabular*}{\textwidth}{@{\extracolsep{\fill}}lccccc@{}}
\hline
& &\multicolumn{4}{c@{}}{$\bolds{x}$} \\[-6pt]
& &\multicolumn{4}{c@{}}{\hrulefill} \\
\multicolumn{2}{@{}l}{$\bolds{l_{\mathrm{\mathbf{ML}}}(\tilde{x}|x)}$} & \textbf{(0, 0)} & \textbf{(0, 1)} & \textbf{(1, 0)} & \textbf{(1, 1)} \\
\hline
{$\tilde{x}$} & (0, 0) & 0 & $\mathrm{FH}$ & $\mathrm{FC
+ FH}$ & $\mathrm{FC}$ \\
& (0, 1) & $\mathrm{FT}$ & 0 & $\mathrm{FC + FT}$ & $\mathrm{FC + FT}$
\\
& (1, 0) & $\mathrm{FC + FT}$ & $\mathrm{FC + FT}$ & 0 & $\mathrm{FT}$
\\
& (1, 1) & $\mathrm{FC}$ & $\mathrm{FC + FH}$ & $\mathrm{FH}$ & 0
\\
\hline
\end{tabular*}
\end{table}

\subsection{Calculating the expected loss under the Markov loss function}

Under the Markov loss function, the expected loss is given by
\[
\mathbb{E}_{\pi(x)} \bigl[ l(\tilde x | x) \bigr] = \sum
_{x} \Biggl[ \sum_{i=1}^{n}
l_M(\tilde{x}_{i} | x_{i}) + \sum
_{i=1}^{n-1} l_T(\tilde {x}_{i, i+1}
| x_{i, i+1}) \Biggr] \pi(x), \label{eq:expected_loss}
\]
where, by exchanging the order of summation,
\begin{eqnarray*}\label{eq:generalmarkovlossfunction}
\mathbb{E}_{\pi(x)} \bigl[ l(\tilde{x}| x) \bigr] &=& \sum
_{i=1}^{n} \sum_{x_{i}}
l_M(\tilde{x}_{i} | x_{i})
\pi(x_{i}) + \sum_{i=1}^{n-1} \sum
_{x_{i, i+1}} l_T(\tilde{x}_{i,i+1} |
x_{i,i+1}) \pi(x_{i, i+1})
\\
&=& \sum_{i=1}^{n} \mathbb{E}_{\pi(x_i)}
\bigl[ l_M(\tilde{x}_{i}; x_{i}) \bigr] +
\sum_{i=1}^{n-1} \mathbb{E}_{\pi(x_{i,i+1})}
\bigl[ l_T(\tilde {x}_{i,i+1} | x_{i,i+1}) \bigr],
\end{eqnarray*}
where $\mathbb{E}_{\pi(x_i)}[l_M(\tilde{x}_i| x_i)]$ and $\mathbb
{E}_{\pi(x_{i,i+1})}[l_T(\tilde{x}_{i,i+1}| x_{i,i+1})]$ are the
expected posterior marginal state and switching losses, respectively.

\subsection{Dynamic programming}

As the expected loss for the Markov loss function is additive, the
prediction $\hat{x}$ that has MEL can be found using the following
dynamic programming recursions (in similar fashion to the Viterbi algorithm):

\subsubsection{Forward recursion}

Compute
\begin{eqnarray*}
\phi_1(k) &=& \min_{j} \gamma \bigl(
\tilde{x}_{1,2} = (j, k) \bigr),
\\
\delta_1(k) &=& \arg\min_{j} \gamma \bigl(
\tilde{x}_{1,2} = (j, k) \bigr),
\end{eqnarray*}
where $k \in\{0, \ldots, S\}$, and then for $i = 2, \ldots, n$,
\begin{eqnarray*}
\phi_i(k) &=& \min_{j} \bigl[
\phi_{i-1}(j) + \gamma \bigl(\tilde{x}_{i,i-1} = (j, k) \bigr)
\bigr],
\\
\delta_i(k) &=& \arg\min_{j} \bigl[
\phi_{i-1}(j) + \gamma \bigl(\tilde {x}_{i,i-1} = (j, k) \bigr)
\bigr],
\end{eqnarray*}
where $\gamma(\tilde{x}_{i,i-1}) = \sum_{x_i} l_M(\tilde{x}_{i}|x_{i})
\pi(x_{i}) + \sum_{x_{i, i-1}} l_T(\tilde{x}_{i,i-1}|x_{i,i-1}) \pi(x_{i,i-1})$.

\subsubsection{Backward trace}

Find $\hat{x}_n = \arg\min_{k} \phi_n(k)$, then $\hat{x}_{i-1} = \delta
_i(\hat{x}_i), i = n-1, \ldots, 2$.

\subsection{Computational requirements}

The order of computation required is ${\cal O}(S^4 N)$, where $S$ is
the number of states and $N$ is the sequence length, since a summation
is required over all possible pairs of the true hidden states $x_{i,
i+1}$ and predictions $\tilde{x}_{i, i+1}$. This can be prohibitive for
applications involving large state spaces but is computationally
manageable for smaller state spaces. In practical situations, though,
it is often the case that the posterior probability distribution
assigns high probabilities to a few transitions while the remainder
have negligible probability. For data exhibiting sparse properties,
these features can be exploited in order to derive approximate
algorithms for inference in Hidden Markov models [see \citet
{Siddiqi2005}] that can offer substantial computational gains at the
expense of little error if the assumption of sparseness holds.

\subsection{Uncertainty in the statistical model}

We have assumed throughout the availability of the exact statistical
model $\pi(x|y)$. In general, of course, it is rare in practice to have
access to the exact statistical model and instead the model is known up
to a form $\pi(x,\theta|y)$ that includes some unknown model parameters
$\theta$. The prediction must then satisfy
\begin{eqnarray*}
\hat{x} &=& \arg\min_{\tilde{x}} \int_{\Theta}
\biggl[ \sum_{ x \in\cal X} l(\tilde{x}| x) \pi(x, \theta|y)
\biggr] \,d\theta
\\
&=& \arg\min_{\tilde{x}} \sum_{ x \in\cal X} l(
\tilde{x}| x) \pi (x|y)
\\
&\approx&\arg\min_{\tilde{x}} \sum_{ x \in\cal X}
l(\tilde{x}| x) \hat{\pi}(x|y),
\end{eqnarray*}
where, in the second line, the independence of the loss function and
the model parameters allow $\theta$ to be integrated out of the model
$\pi(x,\theta|y)$ and the problem is reduced to the same form as
before. The integral required will generally be analytically
intractable and an estimate $\hat{\pi}(x|y)$ must be used that can be
obtained using Monte Carlo simulations, variational methods or by
conditioning on point estimators (such as the MAP).

\subsubsection{Connections to the discrete Fused Lasso method}

Motivated by a similar problem to the one we consider here, \citet
{Zhang2010} adopted a dynamic programming imputation method, based on a
discrete version of the Fused Lasso prior, to penalise state
transitions. The objective function being minimized has the general form
\[
\hat{x} = \arg\min_x \Biggl[ \sum
_{i=1}^n g(x_i;y,\theta) + \lambda \sum
_{i=2}^n (1 - \delta_{x_{t-1},x_t})
\Biggr], \label{eq:generalfusedlasso}
\]
where $g(x_i;y,\theta)$ is a cost term related to data fidelity, for
example, the negative log-likelihood $-\log f(y_i; x_i, \theta)$,
$\lambda$ is a Lasso penalty for state transitions and $\delta
_{x_{i-1},x_i}$ is the Kronecker delta function. Note that \citet
{Zhang2010} actually penalise the absolute difference in signal level
between copy number assignments, but we do not do this here, as, in
contrast to the examination of germline copy number alterations, large
copy number changes in cancers are frequently occurring.

The Fused Lasso term resembles a one-dimensional stationary Markov
Random Field prior of the form $\pi(x) \propto\exp( -\lambda\sum_{i=2}^n (1-\delta_{x_{i-1},x_i}) )$. Since, in one dimension, a
stationary Markov Random Field can be expressed as a Markov chain with
a particular transition matrix [\citet{kesten1976existence}], the method
of \citet{Zhang2010} can be equivalently expressed as finding the
Viterbi sequence for a Hidden Markov model and the Lasso parameter
$\lambda$ provides controls over the prior expected holding times for
the Markov chain. In particular, as only a single parameter is used to
control the state transition penalties, the transition matrix is
symmetric and all states will share the same expected geometric length
distribution. Structured nonsymmetric transitions can be specified by
transition-specific losses.

An illustration of this relationship can be considered in the symmetric
two-state case. The conditional distribution of $X_i | X_{i-1},
X_{i+1}$ for the Markov Random Field is given by
\begin{eqnarray*}
&&\mathrm{Pr}(X_i = x_{i}|X_{i-1} =
x_{i-1}, X_{i+1} = x_{i+1})\\
&&\qquad = \frac{\exp(-\lambda(1-\delta_{x_{i-1},x_i}) ) \exp(-\lambda
(1-\delta_{x_{i},x_{i+1}}) )}{ \sum_{s=1}^S \exp(-\lambda(1-\delta
_{x_{i-1},s})) \exp(-\lambda(1-\delta_{s,x_{i+1}}) ) }.
\end{eqnarray*}

If an equivalent Markov chain exists, with self-transition
probability\break
$\mathrm{Pr}(X_i=x_{i-1}|X_{i-1}=x_{i-1}) = 1-\alpha$, the conditional
distribution can also be expressed as
\begin{eqnarray*}
\mathrm{Pr}(x_{i}|x_{i-1}, x_{i+1}) = \cases{
\displaystyle\frac{\alpha^2}{(1-\alpha)^2 + \alpha^2}, &\quad $x_i \neq x_{i-1}, x_i
\neq x_{i+1} $, \vspace*{2pt}
\cr
\displaystyle\frac{(1-\alpha)^2}{(1-\alpha)^2 + \alpha^2}, &\quad $ x_{i+1}
= x_i = x_{i-1} $, \vspace*{2pt}
\cr
0.5, & \quad$\mbox{otherwise}.$}
\end{eqnarray*}

\begin{figure}

\includegraphics{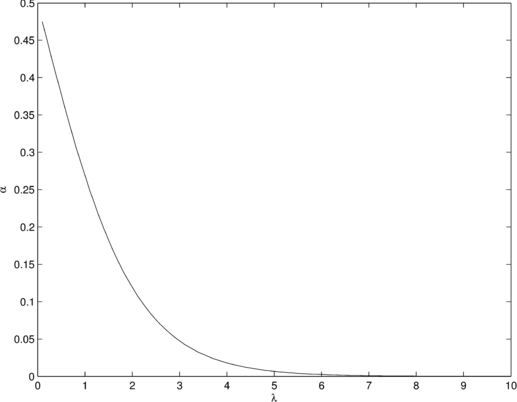}

\caption{The relationship between the Fused Lasso penalty $\lambda$
and the Markov chain transition probability $\alpha$.}
\label{fig:fusedlassohmm}
\end{figure}

By equating these expressions and solving the resulting quadratic, one
can obtain the following relationship between the transition
probability $\alpha$ and the Fused Lasso penalty $\lambda$:
\[
\alpha= \frac{ \beta- \sqrt{\beta}}{ \beta- 1},
\]
where $\beta= \exp(-2\lambda)$. Figure~\ref{fig:fusedlassohmm} shows
that for values of $\lambda$ considered by \citet{Zhang2010} ($\lambda=
0\mbox{--}10$), the transition probability is accordingly small, which is the
desired property for applications in copy number calling applications
where DNA copy number state is expected to persist across sizeable
genomic regions.

As the discrete Fused Lasso of \citet{Zhang2010} implicitly invokes a
structured Hidden Markov model, it therefore can be used as the base
model $\pi(x|y)$ for our decision theoretic approach. In addition,
there are some interesting connections between the discrete Fused Lasso
and our decision theoretic approach. In particular, we can interpret
our method as applying a discrete Fused Lasso type reporting process
\textit{a posteriori} rather than {a priori}. Our method uses
the expected posterior marginal site-wise and pair-wise losses from a
statistical model that has \emph{already} been fitted to data. This
separation of the reporting and model fitting tasks means that our loss
function does not become a proxy for the prior distribution on
sequences. This allows a user to modify the sequence classification
without having to change the statistical model that is fitted to the
data. The benefits of this approach over the Fused Lasso are
illustrated and discussed in the following simulation study.


\section{Results}

\subsection{Simulations}

We performed a simulation study to examine the properties of
predictions made by the use of Viterbi, Fused Lasso and Markov loss
functions in a generic segmental classification setup.

\subsubsection{Assessing performance}

In order to assess performance, we will consider two performance
metrics: (i) the site-wise $(\mathrm{e}_c)$ and (ii) segment-wise $(\mathrm{e}_s)$
misclassification rates. These are defined as
\[
\mathrm{e}_c = 1 - \sum_{i=1}^N
\delta_{\hat{x}_i,x_i}, \qquad  \mathrm{e}_s = 1 - \frac{1}{K} \sum
_{k=1}^K \biggl[ \frac{1}{|S_k|} \sum
_{i \in S_k} \delta_{\hat{x}_i,x_i} \biggr],
\]
where $(\hat{x}_i, x_i)$ are the prediction and true state at the $i$th
position, $K$ is the number of segments in the prediction and $S_k$ is
the subset of locations spanned by the $k$th segment. A segment-wise
misclassification error $e_s = 0$ means all segments are correctly
classified, while $e_s = 1$ means no segments are found correctly. In
this measure, segments contribute \emph{equally} to the segment-wise
performance measure regardless of size. For the motivating application,
this is appropriate, as many small genomic aberrations are of greater
biological importance than larger structural alterations. The latter,
however, contribute more significantly to site-wise classification error.

\subsubsection{Simulation models}

We simulated data sets, each consisting of 100 data sequences of length
$n = 1000$ for four different scenarios. The first two data sets were
simulated according to a Hidden Markov model with Gaussian observation
densities,
\begin{eqnarray*}
\pi \bigl(y_i|x_i = k, \mu, \sigma^2
\bigr) & = & \mathrm{Normal} \bigl(\mu_k, \sigma^2
\bigr),\qquad i = 1, \ldots, n,
\\
\pi(x_i = j |x_{i-1} = k) & = & \Pi_{jk},\qquad i = 2,
\ldots, n,
\\
\pi(x_1 = k) & = & \nu_k,\qquad j = 1, \ldots, S,
\end{eqnarray*}
with a uniform prior state occupancy vector $\nu$ and the transition
matrix $\Pi$ and mean levels $\mu$ are given in Table~\ref{tab:simulations_setup}(a), (b).

\begin{table}
\tabcolsep=0pt
\caption{Parameter settings for simulation study}\label{tab:simulations_setup}
\begin{tabular*}{\textwidth}{@{\extracolsep{\fill}}lccc@{}}
\hline
\textbf{Simulation} & \textbf{Transition matrix,} $\bolds{\Pi}$ & \textbf{State durations,} $\bolds{\lambda}$ &
\textbf{State levels,} $\bolds{\mu}$ \\
\hline
(a) HMM (Sticky) & $ \lleft[
\matrix{
9/10 & 1/30 & 1/30 & 1/30 \vspace*{2pt}\cr
1/30 & 9/10 & 1/30 & 1/30 \vspace*{2pt}\cr
1/30 & 1/30 & 9/10 & 1/30 \vspace*{2pt}\cr
1/30 & 1/30 & 1/30 & 9/10}
\rright] $ & n/a & $\{ -1, 0, 1, 2\}$ \\[24pt]
(b) HMM (Dynamic) & $\lleft[
\matrix{
0.5 & 0.2 & 0.2 & 0.1 \vspace*{2pt}\cr
0.4 & 0.6 & 0.0 & 0.0 \vspace*{2pt}\cr
0.0 & 0.1 & 0.7 & 0.2 \vspace*{2pt}\cr
0.2 & 0.0 & 0.0 & 0.5}
\rright]$ & n/a & $\{ -1, 0, 1, 2\}$ \\[24pt]
(c) HSMM (4-state) & $\lleft[
\matrix{
0.0 & 0.2 & 0.5 & 0.3 \vspace*{2pt}\cr
0.2 & 0.0 & 0.5 & 0.3 \vspace*{2pt}\cr
0.1 & 0.3 & 0.0 & 0.6 \vspace*{2pt}\cr
0.2 & 0.4 & 0.4 & 0.0}
\rright]$ & $\{ 20, 50, 20, 10 \}$ & $\{ -1, 0, 1, 2\}$ \\[24pt]
(d) HSMM (5-state) & $\lleft[
\matrix{
0.0 & 0.25 & 0.25 & 0.25 & 0.25 \vspace*{2pt}\cr
0.25 & 0.0 & 0.25 & 0.25 & 0.25 \vspace*{2pt}\cr
0.25 & 0.25 & 0.0 & 0.25 & 0.25 \vspace*{2pt}\cr
0.25 & 0.25 & 0.25 & 0.0 & 0.25 \vspace*{2pt}\cr
0.25 & 0.25 & 0.25 & 0.25 & 0.0 }
\rright]$ & $\{ 30, 50, 30, 20, 10 \}$ & $\{ -1, 0, 1, 2, 5\}$ \\
\hline
\end{tabular*}
\end{table}

The third and fourth data sets were generated according to Hidden
Semi-Markov model sequences via the following scheme:
\begin{eqnarray*}
y_t | z_t = s, \qquad \sigma^2 &\sim&
\mathrm{Normal} \bigl(\mu_s, \sigma^2 \bigr),
\\
z_{t+1:t+\Delta_i} &=& x_i, t = \sum_{j=1}^{i-1}
\Delta_j,
\\
\Delta_i | x_i = k,\qquad \lambda& \sim&\mathrm{Poisson}(
\lambda_k),\qquad i = 2, \ldots, N,
\\
p( x_i = j | x_{i-1} = k ) &=& \Pi_{jk},\qquad i = 2,
\ldots, N, (j,k) \in\{ 1, \ldots, S\}^2
\\
p( x_1 = j ) &=& \nu_j,\qquad j = 1, \ldots, S,
\end{eqnarray*}
where the transition matrix $\Pi$, state durations $\lambda$ and mean
levels $\mu$ are shown in Table~\ref{tab:simulations_setup}(c), (d).

\subsubsection{Statistical inference}

We fitted a Hidden Markov model with Gaussian observation densities to
each data sequence. We assumed that the parameters of the observational
density $(\mu, \sigma^2)$ are given, but we used a standard
expectation-maximization (or Baum--Welch algorithm) to obtain maximum
likelihood parameter estimates for the prior state occupancy vector
$\hat{\nu}$ and transition matrix,\vadjust{\goodbreak} $\hat{\Pi}$. Note that our primary
interest here is the methods for \emph{reporting} sequence predictions
and not the model fitting procedures themselves which we consider a
separate exercise. Given the parameter estimates, we applied three
methods for segmental classification: (i) we used the Viterbi algorithm
to find the most probable state sequence $\hat{x}_{v}$; (ii) the
discrete Fused Lasso method with a range of penalty values $\lambda=
1\mbox{--}10\mbox{,}000$; and, finally, (iii) we applied the forward--backward
algorithm to obtain the marginal state and switching probabilities $\pi
(x_i|y)$ and $\pi(x_{i,i+1}|y)$ and applied our decision-theoretic
approach with loss parameter values $\mathrm{FC} = 1$, $\mathrm{FN} =
1$ and a range $\mathrm{FT} = 1\mbox{--}10\mbox{,}000$.

\begin{figure}

\includegraphics{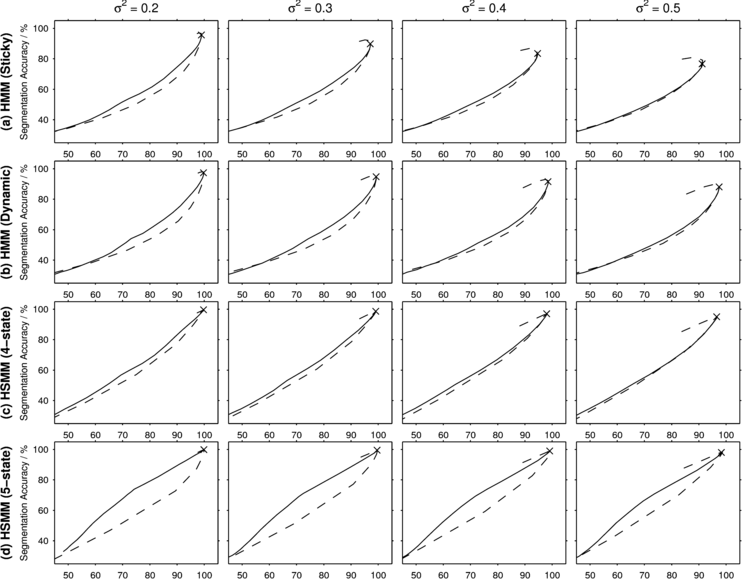}

\caption{Classification of simulated Markov and Semi-Markov sequences
under first-order Markov assumptions. $(-)$ Markov loss. $(--)$ Fused
Lasso. $(\times)$ Viterbi.}\label{fig:summary}
\end{figure}

\subsubsection{Results}

Figure~\ref{fig:summary} shows the average performance of the three
segmentation methods on the four data sets. The Viterbi segmentation
gives excellent site-wise and segment-wise classification accuracy in
all cases. Similar classification performance may be achieved using the
Fused Lasso for a certain choice of penalty parameter $\lambda$. This
parameter would need to be learnt in real applications. For our
decision-theoretic approach, Viterbi-like performance can be achieved
using a default choice of unit loss parameters $\mathrm{FC}=\mathrm
{FH}=\mathrm{FT}=1$ which is convenient for default analyses.

\begin{figure}

\includegraphics{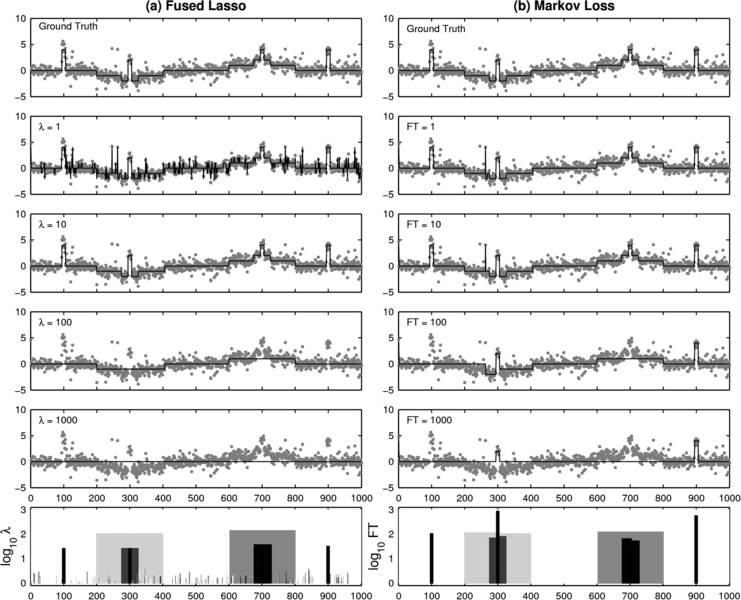}

\caption{Example segmentations using the Fused Lasso and Markov loss
functions for different transition penalties.}\label{fig:example_seg}
\end{figure}

We remark that the Viterbi and Fused Lasso solutions are only available
as we condition on fixed or point parameter estimates. In a full
Bayesian analysis, where Markov chain Monte Carlo (MCMC) methods are
used to sample from the joint posterior distribution, these solutions
would not be available. However, our decision-theoretic approach can
utilise Monte Carlo approximations of the posterior expected marginal
losses and can be applied to MCMC output.

Our principle interest, though, is not the single prediction provided
by Viterbi but the Fused Lasso and our proposed decision-theoretic
approach for exploring alternative segmentations. In this case, by
increasing the transition penalties ($\lambda$ and $\mathrm{FT}$,
resp.), each method is able to produce less complex (smoother)
segmentations with fewer segments. However, Figure~\ref{fig:summary}
shows that for a given site-wise classification accuracy, our
decision-theoretic approach is able to attain a higher segment-wise
classification accuracy than the Fused Lasso method.

Figure~\ref{fig:example_seg} explains the differing segmentation
behaviours. As shown previously, the Fused Lasso penalty $\lambda$ is
related to the prior expected segment length, and large values of
$\lambda$ imply a preference for larger (and therefore fewer) segments.
As a consequence, the short segments tend to be the first to be
eliminated from the Fused Lasso segmentations, while larger segments
are retained. This is because the contribution of small segments to the
overall sequence likelihood is insufficient to justify the penalty of
having two breakpoints to define the small segment.

With our decision-theoretic approach, when computing the expected loss,
the loss penalties are \emph{scaled} by the posterior marginal
site-wise and transition probabilities. Hence, as the penalty on false
transitions $(\mathrm{FT})$ is increased, it is those breakpoints that
are associated with low probability state transitions which are
eliminated first. The segmentations that are produced using the Markov
loss function therefore show a reduction in complexity as the
transition loss $\mathrm{FT}$ is increased, but retain the short, high
signal segments in the data sequence with high probability breakpoints.
We shall see the practical implications of this in the following
application study.

\subsection{Application: DNA copy number profiling of colorectal cancer}

\subsubsection{Setup}

We now consider the use of our methods as an augmented step in existing
Hidden Markov model based approaches for classifying DNA copy number
alterations. One of the problems with such a task is the difficulty of
making formal performance assessments due to a unavailability of ``gold
standard'' genome-wide copy number profiles for cancers. Standard
experimental approaches, such as FISH or PCR, lack the resolution and
throughput necessary to confirm the hundreds to thousands of possible
findings arising from more modern technologies based on microarrays of
next generation sequencing technologies. As a consequence, in the
absence of ground truth data, we adopted the following simulation set
up to produce realistic data sets for evaluation.

\begin{figure}

\includegraphics{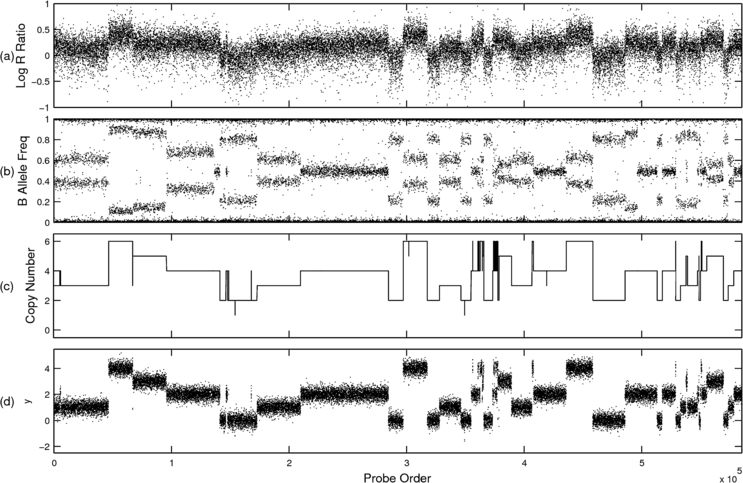}

\caption{Cancer simulation strategy. \textup{(a)}, \textup{(b)} SNP data from the original
colorectal tumor was analysed using OncoSNP [Yau et~al. (\citeyear{Yau2010})] to obtain
a copy number profile \textup{(c)}. Using this copy number profile we simulated
a new data set \textup{(d)} upon which we tested the Viterbi, Fused Lasso and
our decision-theoretic approaches for segmental classification.}
\label{fig:colonexample}
\end{figure}

We collated a genome-wide DNA copy number data set derived from a
recent study of colorectal cancers [\citet{Christie2012}] consisting of
over 630 colorectal tumours. Secondly, for each tumour, the raw
microarray data was processed using a state-of-the-art method, OncoSNP
[\citet{Yau2010}], to infer the DNA copy number profile. Finally, from
this collection of tumour copy number profiles, we then simulated a
series of one-dimensional data sets with Student $t$-distributed noise
using these copy number profiles as a scaffold. The simulation strategy
is illustrated graphically in Figure~\ref{fig:colonexample} using a
data set derived from a colorectal tumour exhibiting chromosomal
instability---a common phenomenon in colorectal cancers. Chromosomal
instability gives rise to large segments with shorter segments
interspersed along the genome residing at sites containing genes with
potential oncogenic or tumor suppressing activity.

This strategy allows us to generate copy number sequence data with
real-world characteristics where we know the truth, and hence better
understand the effect of using Viterbi, the Fused Lasso and our
preferred method based on Markov loss functions for segmental
classification. This partially circumvents the lack of ``gold standard''
copy number profiles for complex tumour samples, without which we are
not able to verify the accuracy of the copy number profile predictions
that would be inferred.

\subsubsection{Simulations}

Given a DNA copy number profile $x_1, \ldots, x_N$ involving $S$ copy
number states, we simulated a data set $y$ according to the following scheme:
%
\begin{equation}
y_i | x_i = k, \qquad\sigma^2 \sim
\mathrm{Student} \bigl(\mu_k, \sigma^2, \nu \bigr),\qquad i =
1, \ldots, N,
\end{equation}
where $\nu= 4$ and $\mu= \log(k/2)$ and $\mu= -4$ for $k=0$ in the
simulations (our simulations mimic the nonlinear response behaviour of
homozygous deletions that involve zero copy number in microarray
experiments). As before, we fitted a Hidden Markov model to the data
using the EM algorithm to obtain maximum likelihood estimates of the
initial state occupancy vector and transition matrix. We applied the
Viterbi algorithm, Fused Lasso and our decision-theoretic method to
give a segmental classification of the data compared to the actual
profile used to generate the data sequence.

\begin{figure}

\includegraphics{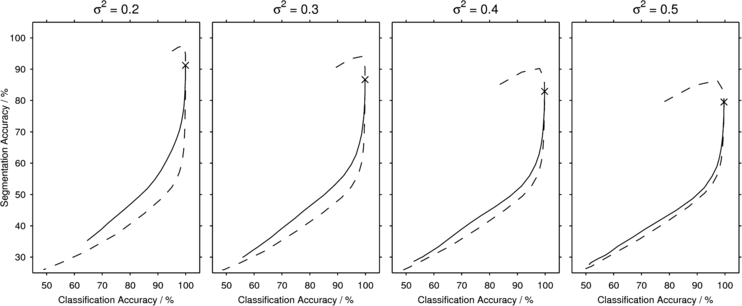}

\caption{Classification performance on the colon cancer data set.
$(-)$ Markov loss. $(--)$ Fused Lasso. $(\times)$ Viterbi.}
\label{fig:colon}
\end{figure}

\begin{figure}

\includegraphics{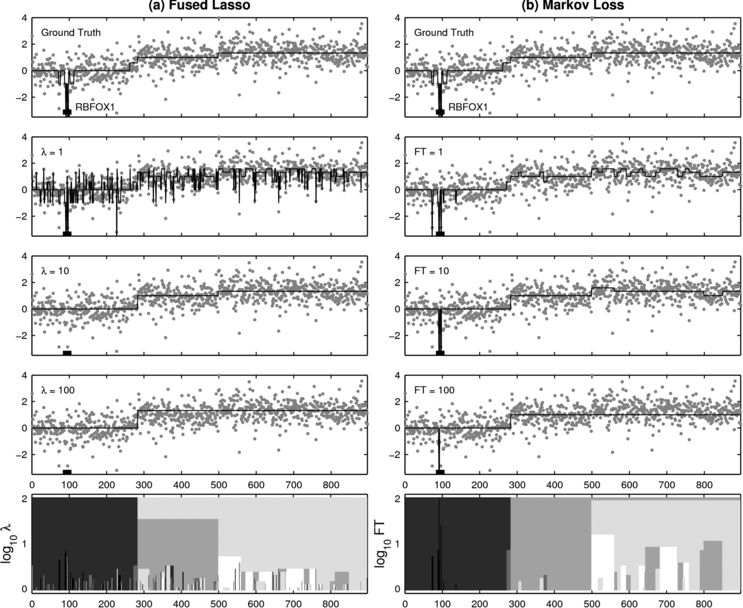}

\caption{Example segmentations using the Fused Lasso and Markov loss
functions for a tumour containing an RBFOX1 deletion.}
\label{fig:rbfox1}
\end{figure}

Note, for these applications, a true physical basis for the statistical
model $\pi(x)$ is unknown and first-order Markov models are often used
as an approximation. Semi-Markov models provide greater modeling
flexibility but are rarely used in genomic applications, as the data
sets involve long sequences (in our CRC application $N = 6 \times
10^5$). Inference methods for the semi-Markov models have computational
requirements that are order ${\cal{O}}(S^2 N^2)$ [\citet
{Murphy02hiddensemi-markov}], which preclude their use in real applications.

\subsubsection{Results}

Figure~\ref{fig:colon} shows that using the Markov loss function we
were able to achieve improved segmental classification rates compared
to the Fused Lasso for the colon cancer data set. A specific example is
illustrated in Figure~\ref{fig:rbfox1} which shows data simulated based
on a tumour carrying a number of large copy number alterations on
chromosome 16 and a small homozygous deletion involving the alternative
splicing factor RBFOX1. Deletions of RBFOX1 are a recurrent event in
colorectal cancer [\citet{Network2012}] and were recently found to have
high prevalence in patients from a Bangladeshi population versus
Caucasians [\citet{Sengupta2013}]. Deletions in this region are complex,
with focal deletions targeting the 5$'$ end of the gene, and have been
shown to affect mRNA and protein expression in colorectal cell lines
and tumours. A copy number profile of this tumour should ideally report
the presence of the RBFOX1 deletion, but the other larger copy number
changes may be of less importance as they are likely to be passenger
events formed due to genomic instability during tumour
evolution.\looseness=-1

In the Fused Lasso segmentations, we can encourage smoother
segmentations by increasing the transition penalty. However, the effect
of using larger penalties causes\vadjust{\goodbreak}  the important RBFOX1 deletion to be
eliminated and only the larger copy number alterations are retained.
With our decision-theoretic approach, the RBFOX1 deletion is identified
even when the false transition loss parameter was increased---we are
able to achieve smoothing without losing this important fine detail.

These results indicate that our method could be used to augment
existing Hidden Markov model-based calling algorithms for copy number
aberrations, such as those by \citet{Sun2009,Yau2010} and \citet{Li2011},
with a sequence classification algorithm that provides a more flexible
alternative to the Viterbi algorithm and has improved segmental
classification performance relative to the Fused Lasso method. In
particular, we demonstrate the adaptive nature of the Markov loss
function, in terms of its ability to provide reduced complexity copy
number segmentations while retaining important features such as small
homozygous deletions or gene amplifications. This may assist cancer
researchers in isolating important genetic alterations of interest in
cases where a default Viterbi segmentation might produce unmanageably
complex copy profiles.


\section{Discussion}

Segmental classification problems are ubiquitous across many fields,
including signal processing, finance and, more recently, genomics. We
have introduced a Markov loss function that allows a user to take their
preferred statistical model $\pi(x)$ of the sequence $x$ and obtain a
sequence prediction $\hat{x}$ whose properties can be adjusted in an
intuitive way by specifying loss parameters on state and transition
errors. The calculation of the posterior expected loss with respect to
a Markov loss function was shown to have a simple form and a dynamic
programming algorithm was provided to compute the state sequence with
the minimum expected loss.

Although the emphasis in this presentation was on the Hidden Markov
model as the statistical model $\pi(x)$, this method can be applied to
any statistical model for the segmentation and classification of linear
sequence data that can provide estimates of the marginal state
transition probability $\pi(x_{i,i+1})$. Therefore, it can be used to
augment, without modification, many existing statistical methods for
analyzing sequence data, such as those based on semi-Markov models,
change point methods [\citet{Fearnhead2007}] or product partition models
[\citet{Barry1992}]. While it is a relatively simple addition, the
application of this method could greatly enhance the adaptability of
many existing statistical algorithms, transferring power to the
experimenter to allow them to assign losses to various error types
relevant to their own study.

Our approach can be considered to be a specific form of the loss
functions considered by \citet{rue1995new} in Bayesian imaging
applications. \citet{rue1995new} considered a more complex
two-dimensional domain, using Markov Random Field priors, where exact
enumeration of the optimal decision is impossible and numerical
optimisation using computationally-intensive MCMC and simulated
annealing is required. Recently, \citet{lember2010generalized} have also
considered generalised risk-based inference for Hidden Markov models
including a subclass of posterior decoding schemes that can be viewed
as hybrids of the Viterbi and marginal approaches.

Throughout this paper we have not explicitly stated how the loss values
should be selected. This is purposeful because the selection of the
costs associated with various error types is \textit{study-dependent} and
the individual data analyst must balance the appropriate losses for the
particular application. For example, in genomics, costs might be
related to tangible quantities such as the financial, time and manpower
requirements for follow-up studies and validation taken upon the
predictions. We indicate that a default choice of loss parameters can
lead to a Viterbi-like performance.

It is also of further research interest to characterise the effect on
predictions when only an approximation of the statistical model is
available. Furthermore, in some applications there may be some utility
in combining of Markov loss functions on the hidden state sequence $x$
and loss functions on the model parameters $\theta$. The Markov loss
function introduced here focuses on costs associated with
classification errors of the hidden state sequence and assumes that the
model parameters are in some sense nuisance variables. There are
applications where both the state sequence and model parameters may be
of interest; for example, the transition matrix may have some
interpretation for a given application and a loss function may be given
on $\theta$. In these instances it may be necessary to derive optimal
joint predictions $(\hat{x}, \hat{\theta})$ under the appropriate loss
functions.

\section*{Acknowledgements}
We thank Doctor Oliver Sieber and Professor Andrew Silver for the
colorectal cancer genotyping data. We also thank Professors Peter
Green, Peter Donnelly and Havard Rue, Doctors Juri Lember and Alexey
Kolydenko, the Associate Editor and the two anonymous referees whose
comments greatly aided in improving this work.

%

%



\printaddresses

\end{document}